\documentclass[letterpaper,twocolumn,10pt]{article}
\usepackage[letterpaper,margin=0.75in,columnsep=0.25in]{geometry}
\usepackage{times}
\usepackage{graphicx}
\usepackage{natbib}
\usepackage{caption}
\usepackage{booktabs}
\usepackage{amsmath,amssymb}
\usepackage{algorithm}
\usepackage{algpseudocode}
\usepackage{xcolor}
\usepackage{microtype}
\usepackage{float}
\usepackage{tikz}
\usepackage{pgfplots}
\pgfplotsset{compat=1.17}
\usepackage{placeins}
\usepackage{balance}
\usepackage[hyphens]{url}
\usepackage[colorlinks,citecolor=blue,linkcolor=blue,urlcolor=blue]{hyperref}
\algrenewcommand\algorithmicrequire{\textbf{Input:}}
\algrenewcommand\algorithmicensure{\textbf{Output:}}
\makeatletter
\def\fps@figure{htbp}
\def\fps@table{htbp}
\makeatother
\usepackage{authblk}
\usepackage{comment}

\title{\bf Sign-hedged Beam QAOA:\\
Scaling Quantum Optimization Algorithms}

\author{Anupama Ray \thanks{Corresponding author: anupamar@in.ibm.com}}
 \author{Arush Sharma\thanks{Arush worked on this problem during his summer internship at IBM Research.}}
\affil{IBM Research}

\date{}

\begin{document}
\maketitle

\begin{abstract}

Quantum Approximate Optimization Algorithm (QAOA) encodes a combinatorial objective as an Ising Hamiltonian and alternates cost and mixer operations for $p$ rounds. Recursive QAOA (RQAOA) scales further by using the resulting correlations—whether variables tend to agree or disagree—to eliminate variables one at a time. But each inferred sign is accepted irrevocably, so a single early mistake propagates through every later reduction and can discard the global optimum. As elimination proceeds, residuals shrink while effective edge density typically rises, so later decisions face denser Ising problems—a regime that fully connected Sherrington–Kirkpatrick (SK) instances stress-
test directly. 
This paper proposes sign-hedged beam RQAOA, which retains the variable or pair selected by standard RQAOA, explores both possible sign assignments, keeps the best $B$ residual problems according to a classical score, and completes each survivor with ordinary RQAOA.
On frustrated SK $\pm1$ instances, where standard recursion collapses to AR $0.474$, beam recovers near-optimal AR $0.998$ at the same $p$. We evaluate the workflow on 50+ instances from HamLib and QOBLIB, spanning 4-/5-regular and Erdős–Rényi MaxCut graphs, Sherrington–Kirkpatrick (SK) spin glasses, and cardinality-constrained portfolio optimization, with up to 60 qubits and hardware runs on \texttt{ibm\_kingston} and \texttt{ibm\_marrakesh}. Matched simulator and device cells show consistent beam rankings: simulation and hardware agree that beam recovers near-optimal approximation ratio (AR) where standard recursion fails. These gains come from additional residual-circuit evaluations, classical pruning, and exact solution of a small core, rather than from increasing $p$: beam is a decision-side optimization atop fixed-$p$ QAOA/RQAOA, not a claim that quantum circuits alone scale with qubit count.
\end{abstract}

\section{Introduction}
\label{sec:intro}

The Quantum Approximate Optimization Algorithm (QAOA)~\citep{farhi2014qaoa} is one of the most extensively studied gate-based quantum algorithms for combinatorial optimization problems (COPs). QAOA encodes the cost function into a problem Hamiltonian and alternates layers of parametric cost and mixer unitaries to amplify high-quality solutions. It enjoys a performance guarantee that, in the infinite-depth limit, converges to the optimum---a property inherited from its connection to quantum adiabatic evolution~\citep{farhi2000adiabatic}. Despite its theoretical appeal and compatibility with near-term hardware, QAOA faces several practical limitations. First, the classical outer loop that tunes the variational parameters $\gamma$ and $\beta$ faces a rugged landscape with many local minima~\citep{bittel2021nphard,zhou2020qaoa}. Second, barren plateaus---vanishing gradients in the cost landscape---can make gradient-based training intractable at larger scales~\citep{cerezo2021barren,larocca2022barren}. Third, deep QAOA circuits, while more expressive, are often infeasible on noisy intermediate-scale quantum (NISQ) devices, where gate errors and decoherence limit useful depth~\citep{harrigan2021qaoa,shaydulin2024scaling}. A range of strategies address these challenges. Fixed-parameter protocols such as linear-ramp schedules (LR-QAOA) remove instance-specific angle search~\citep{montanez2024lr,brandao2018concentration}, yet
 the correlations obtained at modest $p$ can remain unreliable on difficult, frustrated graphs. Parameter-transfer and related multi-objective QAOA approaches reuse trained angles across instances~\citep{kale2025qaoa}. Recursive QAOA (RQAOA) iteratively reduces problem size by fixing variables from quantum correlators, enabling fixed-$p$ circuits to tackle larger problems~\citep{bravyi2020obstacles,bravyi2022rqaoa}. Warm-starting strategies bias the search with classically obtained solutions~\citep{egger2021warm}. Collectively, these advances aim to keep useful $p$ modest while reducing the optimization burden. Concurrent and independent work by Hadfield et al. ~\citep{hadfield2026iterative} reframes recursive reduction as iterative quantum optimization and introduces Bound-Aware Tree-IQO, which replaces RQAOA's single irreversible path with a reversible branch-and-bound tree that backtracks over quantum-selected reductions and prunes with classical bounds. Our sign-hedged beam is a lightweight, hardware-run instance of the same reversibility principle, bounded to beam width B at fixed p rather than full tree search.

Combinatorial objectives such as MaxCut ask for an assignment of classical spins $z_i\in\{-1,+1\}$ that extremizes an Ising energy~\citep{lucas2014ising,goemans1995gw,sherrington1975sk},
\begin{equation}
E(z)=-\sum_{i<j}J_{ij}z_i z_j-\sum_i h_i z_i.
\label{eq:ising-intro}
\end{equation}
MaxCut divides the vertices of a graph into two groups and rewards edges whose endpoints lie in different groups. The spin $z_i$ records which group contains vertex $i$. With an appropriate choice of $J_{ij}$ and an irrelevant constant shift, minimizing the Ising energy is equivalent to maximizing the cut weight.
QAOA turns this into a layered quantum circuit. It starts from $\lvert+\rangle^{\otimes n}$ (equal superposition over all $2^n$ bitstrings) and applies $p$ \emph{rounds}. Each round has two unitaries: a \emph{cost} layer $U_C(\gamma_\ell)=\exp(-i\gamma_\ell H_C)$ that imprints problem-dependent phases from the Ising Hamiltonian $H_C$, then a \emph{mixer} $U_M(\beta_\ell)=\exp(-i\beta_\ell\sum_i X_i)$ that redistributes amplitude among bitstrings. Stacking $p$ rounds and measuring yields
\begin{equation}
\lvert\psi_p(\beta,\gamma)\rangle
=\prod_{\ell=1}^{p}U_M(\beta_\ell)\,U_C(\gamma_\ell)\,\lvert+\rangle^{\otimes n}.
\label{eq:qaoa-state}
\end{equation}
Thus $p$ is the number of logical QAOA rounds (QAOA layer depth). Compiled physical gate depth also depends on the graph, gate decomposition, qubit mapping, and device. The continuous parameters $(\beta_\ell,\gamma_\ell)$---$2p$ angles---are usually found by a classical optimizer that repeatedly queries the quantum device~\citep{cerezo2021vqa,zhou2020qaoa}. Deeper $p$ can build stronger correlations, but enlarges that search and the compiled noise footprint~\citep{preskill2018nisq,harrigan2021qaoa}.

In this paper we pursue two complementary directions toward a stronger fixed-$p$ optimizer. First, we use an evolutionary CliffSearch~\citep{cliffsearch} workflow to discover a better fixed-angle schedule for QAOA, avoiding instance-wise iterative angle updates. Second, we develop sign-hedged beam RQAOA as a decision-side improvement to recursive elimination. Fixed schedules improve state preparation, while sign hedging targets the irreversible recursive decisions that follow. A structural point motivates the second lever: as RQAOA eliminates variables, the residual graph shrinks while effective edge density typically increases, so later correlators are estimated on denser Ising problems. 
Fully connected SK instances already live in that densified regime, and recovering near-optimal approximation ratio (AR) there with beam---while standard RQAOA collapses---is evidence that the decision policy can handle the densification that recursive reduction induces on larger graphs, provided residual QAOA circuits remain executable. Beam remains an optimization \emph{on top of} fixed-$p$ QAOA/RQAOA rather than a claim that quantum circuits alone scale with qubit count. Together, the evolved power-law schedule (``gold'' in this paper) and sign-hedged beam RQAOA form a full methodology in which gold supplies the parameters and beam supplies robustness when correlator signs are ambiguous. We evaluate primarily at $p{=}2$; the schedule-transfer study and selected simulation cells use other $p$. The key contributions of this paper are:
\begin{enumerate}
    \item Sign-hedged beam RQAOA: a bounded decision-side search that hedges elimination signs at fixed $p$, recovering near-optimal solutions where irreversible standard RQAOA fails on densified residuals (notably SK $\pm1$).
    \item An evolved fixed-angle schedule (``gold'') with the same interface as LR-QAOA, improving transfer success probability and selected plain/recursive AR comparisons without instance-wise angle search.
    \item Exhaustive experimentation on both simulator and hardware across multiple combinatorial optimization problems (MaxCut, SK, portfolio), showing consistent gains for the proposed method relative to vanilla QAOA, RQAOA, LR-QAOA, and related baselines. Beam method raises SK $\pm1$ AR from $0.474/0.470$ to $0.998$ ($\approx 2.1\times$) and matches the optimum on ibm\_marrakesh (AR $1.000$). On Erdős–Rényi (Maxcut) beam reaches AR $0.990$ in both MPS simulation and on ibm\_kingston; and for Portfolio (first16) instances Beam ties AR $1.000$ in simulation and hardware.
\end{enumerate}
Section~\ref{sec:cliff} describes the evolutionary schedule discovery and its output; Section~\ref{sec:method} details sign-hedged beam RQAOA. Section~\ref{sec:evaluation} presents the details of graphs chosen from existing benchmark dataset, algorithm pseudocode, experimental protocols and results. 
\section{Evolutionary discovery of fixed parameter schedules}
\label{sec:cliff}

We use two complementary scores. \emph{Success probability} is the Born probability of sampling an optimal bitstring in one shot. \emph{Approximation ratio} (AR) for MaxCut is the best returned cut divided by the certified optimum (portfolio uses the exact feasible objective range). Success can be low while AR is still high if a near-optimal string is returned; the CliffSearch evaluation below reports success, while the beam study reports AR.

\begin{figure}[H]
\centering
\begin{tikzpicture}
\begin{axis}[
  width=\columnwidth,
  height=0.22\textheight,
  xmode=log,
  ymode=log,
  xmin=7, xmax=140,
  ymin=0.1, ymax=1.05,
  xtick={10,100},
  xticklabels={10,100},
  xlabel={QAOA depth $p$},
  ylabel={Success probability},
  title={$n{=}24$ 3-regular MaxCut: gold schedule vs.\ LR-QAOA},
  title style={font=\footnotesize\bfseries},
  legend style={font=\scriptsize, at={(0.03,0.97)}, anchor=north west},
  grid=both,
  grid style={dashed,gray!40},
  tick label style={font=\scriptsize},
  label style={font=\scriptsize},
]
\addplot[color=blue,mark=*,thick] coordinates {(10,0.280) (100,0.952)};
\addlegendentry{CliffSearch gold schedule}
\addplot[color=orange,mark=*,thick] coordinates {(10,0.154) (100,0.901)};
\addlegendentry{LR-QAOA linear schedule}
\end{axis}
\end{tikzpicture}
\caption{Gold schedule vs.\ LR-QAOA on a held-out $n{=}24$ 3-regular MaxCut graph instance. Success probability is the Born weight of an optimal string. Gold raises success from $0.154$ to $0.280$ at $p{=}10$ ($1.82\times$) and from $0.901$ to $0.952$ at $p{=}100$. This is a deep non-recursive (plain) QAOA transfer study, distinct from the mostly $p{=}2$ beam evaluation.}
\label{fig:gold-transfer}
\end{figure}
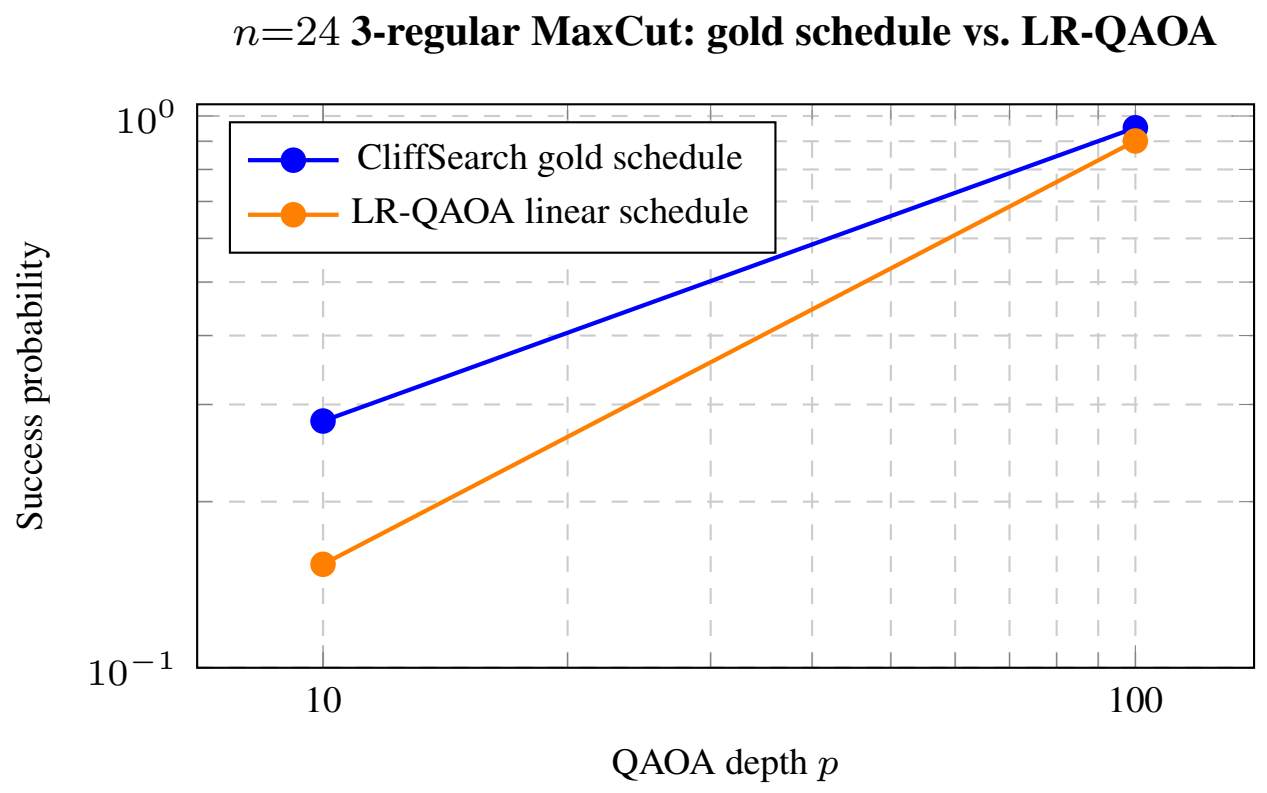

Linear-ramp QAOA (LR-QAOA) replaces instance-wise angle search with a fixed schedule~\citep{montanez2024lr}: a predetermined formula for $(\beta_\ell,\gamma_\ell)$ so every residual uses the same parameters. Here ``LR'' is the original linear ramp; ``gold'' denotes the evolved power-law schedule with the same interface,
\begin{equation}
\beta_\ell=\Delta_\beta\bigl[1-(\ell/p)^{1.65}\bigr],\qquad
\gamma_\ell=\Delta_\gamma\bigl[((\ell+1)/p)^{0.725}\bigr].
\label{eq:gold}
\end{equation}
CliffSearch~\citep{cliffsearch} searched schedule \emph{formulas}, and upon multiple evolutions the strongest emerging schedule is called as gold schedule in the rest of the paper. Gold schedule \ref{eq:gold} is one fixed ramp plugged into the same QAOA runner as LR-QAOA rather than a new algorithm class. On a held-out $n{=}24$ 3-regular MaxCut graph instance, gold raises success probability by $+82\%$ at $p{=}10$ ($0.154\to0.280$, $1.82\times$) and by $+5.6\%$ at $p{=}100$ ($0.901\to 0.952$; as shown in Figure.~\ref{fig:gold-transfer}).  
In additional non-recursive plain AR checks on 4-regular MaxCut, gold is ahead of LR at several higher $p$: at $n{=}40$, gold reaches $1.000$ vs.\ LR $0.981$/$0.990$/$0.990$ at $p{=}4/5/8$; at $n{=}60$, gold leads at $p{=}2$ ($0.824$ vs.\ $0.810$), $p{=}4$ ($0.935$ vs.\ $0.928$), and $p{=}5$ ($0.961$ vs.\ $0.941$) as shown in table \ref{tab:gold-ar}.

\begin{table}[h]
\centering
\setlength{\tabcolsep}{6pt}
\caption{Non-recursive (plain) AR on 4-regular MaxCut (HamLib): evolved
gold schedule vs.\ LR-QAOA at matched $p$, both fixed-schedule (no
per-instance angle search). Bold marks the higher AR in each row.}
\label{tab:gold-ar}
\small
\begin{tabular}{@{}rrrr@{}}
\toprule
$n$ & $p$ & gold & LR \\
\midrule
40 & 4 & \textbf{1.000} & 0.981 \\
40 & 5 & \textbf{1.000} & 0.990 \\
40 & 8 & \textbf{1.000} & 0.990 \\
60 & 2 & \textbf{0.824} & 0.810 \\
60 & 4 & \textbf{0.935} & 0.928 \\
60 & 5 & \textbf{0.961} & 0.941 \\
\bottomrule
\end{tabular}
\end{table}
These are selected gold${>}$LR comparisons rather than an aggregate claim that gold dominates every $p$, and LR still wins some shallow plain settings (e.g.\ $n{=}40$, $p{=}2$). Better schedules improve preparation, but they neither remove the need for larger $p$ when correlations demand it nor protect a recursive solver from a wrong elimination sign. Warm starts, mixers, and counterdiabatic variants likewise attack preparation~\citep{sack2021quantum,egger2021warm,hadfield2019qaoa,wurtz2021counterdiabatic}.

\section{Sign-hedged beam RQAOA}
\label{sec:method}

Recursive QAOA (RQAOA)~\citep{bravyi2020obstacles,bravyi2022rqaoa} shrinks the problem instead of deepening the circuit. After a fixed-$p$ QAOA state, it estimates one-body biases $\langle Z_i\rangle$ (which value spin $i$ prefers) and two-body correlators $\langle Z_i Z_j\rangle$ (positive $\approx$ align, negative $\approx$ anti-align), typically by averaging over shots or from an exact statevector. It then \emph{eliminates}: pick the correlator of largest magnitude, commit to its sign, substitute into equation ~\eqref{eq:ising-intro}, and obtain a smaller \emph{residual} Ising problem. The procedure repeats until a tiny core remains, which is solved exactly before backtracking. As variables are eliminated, residual graphs typically become denser even as $n$ falls, so late-stage decisions are made on harder, more connected Ising problems. Magnitude selects which relation looks informative, while sign selects how to freeze it; because that sign is irreversible, later steps never undo the substitution and one early error rewrites every later residual. Concretely, if RQAOA substitutes $z_i=-z_j$, every later residual assumes that relation is correct, and if the best global solution instead requires $z_i=z_j$, the correct solution has already been removed from the remaining search space. On frustrated SK $\pm1$ graphs (fully connected, thus a densification stress test), many couplings conflict, local correlators can look strong while the global assignment is hard, and largest-$|\mu|$ signs are especially misleading.

\begin{figure*}[t]
\centering
\includegraphics[width=\textwidth,keepaspectratio]{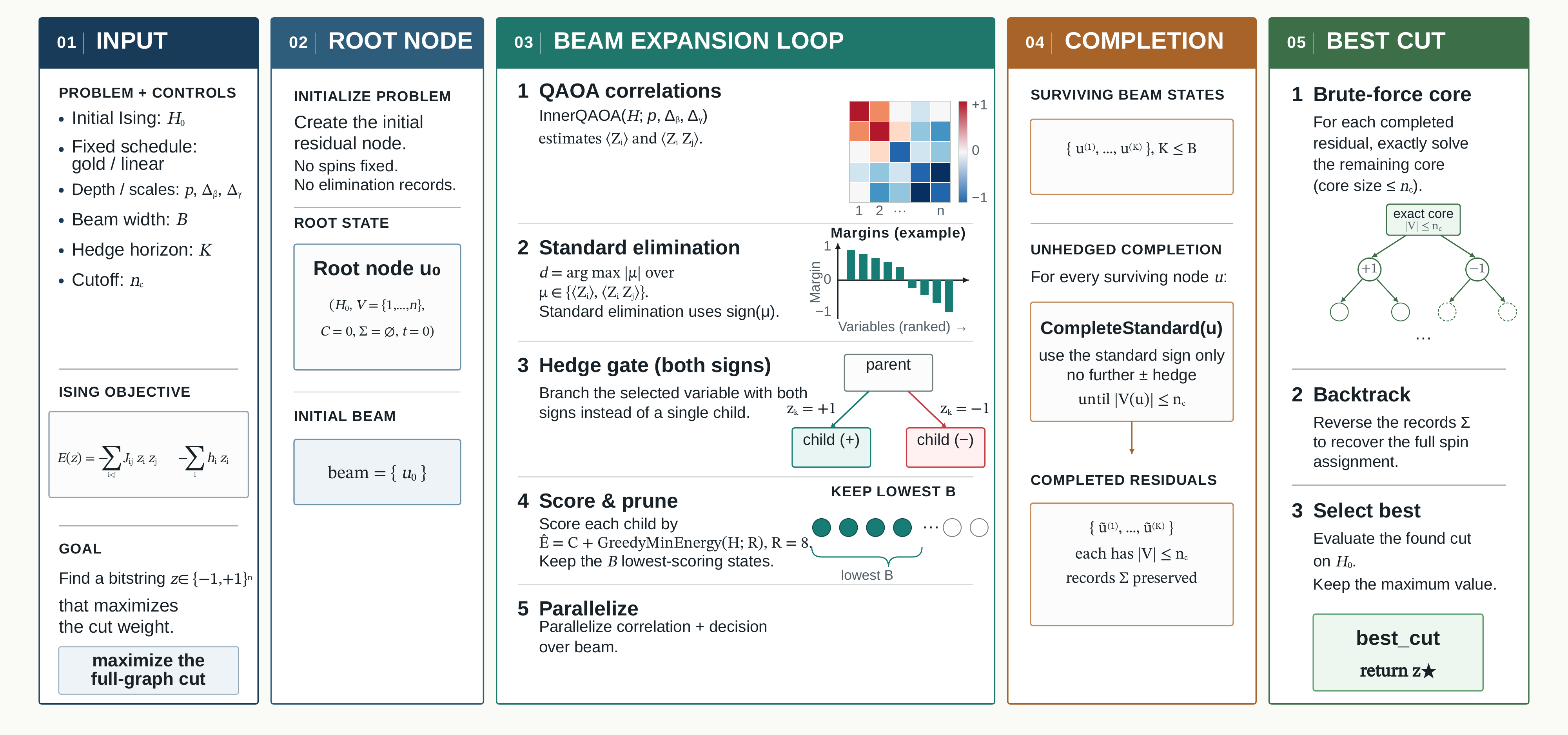}
\caption{Sign-hedged beam RQAOA overview: correlate $\rightarrow$ select $\rightarrow$ hedge signs $\rightarrow$ prune to beam width $B$ $\rightarrow$ complete $\rightarrow$ best cut.}
\label{fig:method}
\end{figure*}

Figure~\ref{fig:method} summarizes the pipeline we study. Non-recursive QAOA (vanilla QAOA) runs one circuit and reads a bitstring; standard RQAOA repeatedly correlates, eliminates, and shrinks along a single sign path; and sign-hedged beam keeps the same selected variable or pair but tries both signs, scores residuals classically, keeps at most $B{=}4$ survivors, and completes each with ordinary RQAOA at fixed $p{=}2$~\citep{lowerre1976beam,harvey1995lds}. Fixed schedules buy better correlators, while beam buys robustness when correlators remain ambiguous, so the novelty is a decision-side policy rather than another ansatz. Experiments on \texttt{ibm\_kingston} and \texttt{ibm\_marrakesh} test whether the policy survives approximate correlators and device noise~\citep{harrigan2021qaoa}. The parameter choices of $B{=}4$ or $p{=}2$ is a selection committed to based on results of multiple ablation studies which consider the result quality, 2-qubit depth post transpilation of circuits on hardware. 

\subsection{From one elimination path to a bounded search}
Algorithm~\ref{alg:beam} gives the corresponding pseudocode. 
Starting from~\eqref{eq:ising-intro}, for a residual Hamiltonian $H$ on surviving spins $V$, \textsc{InnerQAOA} prepares a depth-$p$ fixed-schedule state and estimates $\langle Z_i\rangle$ and $\langle Z_iZ_j\rangle$. An exact statevector stores all quantum amplitudes and introduces no measurement-sampling error; shot-based evaluation estimates observables from repeated measurements; and a matrix-product state (MPS) stores a compressed approximation whose bond dimension controls an accuracy--cost tradeoff~\citep{vidal2003efficient,cirac2021mps}. Small runs use the full quantum state on exact-statevector, while larger MaxCut graphs use MPS correlators from a modest number of shots (approximate, but the same observables the elimination rule consumes). For quantum hardware runs, we use similar number of shots for baseline comparison and also try with multiple levels of increasing shots. 

Standard elimination selects
\[
d=\arg\max_{\mu\in\{\langle Z_i\rangle,\langle Z_iZ_j\rangle\}}|\mu|
\]
and imposes $s=\operatorname{sign}(\mu)$. If $d$ is a single spin, the sign pins $z_d=\pm1$; if $d$ is a pair, it enforces $z_i=\pm z_j$. Substituting into the Ising energy removes that degree of freedom and yields a new residual, which is why the choice is irreversible. Sign hedging leaves $d$ unchanged and generates both $\sigma=+1$ and $\sigma=-1$ children, so subsequent standard eliminations can diverge across branches.

Each beam node stores its residual $H$, alive set $V$, accumulated constant $C$ (energy already fixed by past eliminations), elimination record $\Sigma$, and step $t$. The method parameters are beam width $B$ (maximum number of surviving residual branches), hedge horizon $K$ (number of initial steps that force both signs), restart count $R$ for residual scoring, core size $n_c$ for the exact solve, and an optional margin $\tau$ that continues hedging after step $K$ when the two largest pair-correlation magnitudes are closer than $\tau$ (a closeness / ranking ambiguity of the leading pairs, not a direct estimate of sign uncertainty). For the first $K$ steps both signs are tried. Each child is scored by a cheap classical estimate of residual energy-to-go,
\[
\widehat E=C+\textsc{GreedyMinEnergy}(H;R)
\]
with $R{=}8$ random-restart greedy flips on the residual Ising (a classical lower-energy probe rather than another QAOA call). Lower $\widehat E$ is more promising under minimization, so only the $B{=}4$ best children survive. Survivors then complete without hedging until $n_c$ spins remain; the core is solved exactly, $\Sigma$ is backtracked to rebuild the full assignment, and the cut of that assignment on $H_0$ is scored. For larger MaxCut graphs and the run on \texttt{ibm\_kingston} we use a hedge horizon of $2$ steps---the same role as $K$, but shorter than the schematic $K{=}4$ in Algorithm~\ref{alg:beam}---with $\tau{=}0.10$ (Erd\H{o}s--R\'enyi $k{=}2$) and $\tau{=}0.15$ (5-regular), and cores of size $n_c{=}15$. These are empirical settings rather than a universal $K{=}4$ claim. After hedging, every survivor follows unhedged \textit{CompleteStandard}.

\subsection{Cost and limiting cases}
Sign hedging keeps the circuit depth $p$ fixed and never deepens the quantum circuit. Every added cost is classical. First, because each step can branch into two children, a single beam step issues up to $B$ residual-circuit evaluations instead of one. Second, each child is scored by a cheap classical greedy search over the residual Ising problem, which never touches the quantum device. Third, once the beam reaches the cutoff size, the small final core is solved exactly on a classical computer. None of these three costs changes how deep the quantum circuit has to be, thus being entirely classical cost. 

This distinction matters because circuit depth is the quantity that limits what runs reliably on today's hardware: deeper circuits accumulate more gate error and decoherence before measurement. Beam leaves that quantity fixed at the chosen $p$ and spends only classical effort---more circuit evaluations, more greedy scoring, and one exact solution of a small final core---to recover quality. The additional evaluations are also parallelizable: within a single step, the up-to-$B$ \textsc{InnerQAOA} calls are independent of one another and can be issued at the same time, while the scoring step that selects survivors runs serially afterward.

Two limiting cases bound the method. Setting $B{=}1$ with no hedging keeps a single branch at every step and recovers standard fixed-schedule RQAOA exactly, so beam is a strict generalization of the recursive baseline rather than a different algorithm. Increasing $B$ widens the search but with diminishing returns: we use $B{=}4$ as the default under serial cost, and wider beams ($B{=}6/8$) do not improve AR monotonically and are reported only as ablations (Table~\ref{tab:wall}).

\begin{figure*}[t]
\small
\begin{algorithmic}[1]
\Require $H_0$; fixed schedule $(p,\Delta_\beta,\Delta_\gamma)$; beam width $B$; hedge horizon $K$; cutoff $n_c$; optional margin $\tau$; restarts $R$
\Ensure Full assignment $z^\star$ with the largest found cut value on $H_0$
\State $\mathrm{beam}\gets\{(H_0,V{=}\{1,\ldots,n\},C{=}0,\Sigma{=}\emptyset,t{=}0)\}$
\While{$\mathrm{beam}\neq\emptyset$ \textbf{and} $|V(u_{\mathrm{first}})|>n_c$}
  \If{$\tau$ is unused \textbf{and} $t(u_{\mathrm{first}})\ge K$}
    \State \textbf{break}\Comment{leave the hedging phase}
  \EndIf
  \State $\mathrm{children}\gets\emptyset$
  \ForAll{$u=(H,V,C,\Sigma,t)\in\mathrm{beam}$}
    \State $\{\langle Z_i\rangle,\langle Z_iZ_j\rangle\}\gets\mathrm{InnerQAOA}(H;p,\Delta_\beta,\Delta_\gamma)$
    \State $d\gets\mathrm{StandardEliminate}(\cdot)$;\quad $s\gets\operatorname{sign}(\mu)$;\quad $\mathrm{hedge}\gets(t<K)$
    \If{$\tau$ is set \textbf{and} $\neg\mathrm{hedge}$}
      \State let $|c_{(1)}|,|c_{(2)}|$ be the two largest pair-correlation magnitudes
      \State $\mathrm{hedge}\gets(|c_{(1)}|-|c_{(2)}|<\tau)$
    \EndIf
    \If{$\mathrm{hedge}$}
      \State add $\mathrm{Reduce}(u,d,+1)$ and $\mathrm{Reduce}(u,d,-1)$ to $\mathrm{children}$
    \Else
      \State add $\mathrm{Reduce}(u,d,s)$ to $\mathrm{children}$
    \EndIf
  \EndFor
  \State score children by $\widehat E\gets C+\mathrm{GreedyMinEnergy}(H;R)$; retain the lowest-scoring $B$
\EndWhile
\ForAll{$u\in\mathrm{beam}$}
  \State $u\gets\mathrm{CompleteStandard}(u)$;\quad $z\gets\mathrm{Backtrack}(\Sigma,\mathrm{ExactSolve}(u.H))$;\quad retain $z$ if best on $H_0$
\EndFor
\State \Return $z^\star$
\end{algorithmic}
\captionsetup{type=algorithm}
\caption{Sign-hedged beam RQAOA (defaults $B{=}4$, $K{=}4$, $R{=}8$, $n_c{=}5$; larger MaxCut / hardware: hedge horizon $2$, $n_c{=}15$).}
\label{alg:beam}
\end{figure*}

\section{Evaluation}
\label{sec:evaluation}

\subsection{Protocol and baselines}

We evaluate four families across different $p$, with most reported beam results at $p{=}2$ unless noted: HamLib MaxCut~\citep{sawaya2024hamlib} (Erd\H{o}s--R\'enyi $k{=}2$/$k{=}3$ and 4-/5-regular graphs); Sherrington--Kirkpatrick (SK) $\pm1$ and Gaussian spin glasses; and cardinality-constrained portfolio optimization built from a QOBLIB~\citep{qoblib2024} 50-asset covariance file by taking the first-$n$ assets (\texttt{first16}/\texttt{first20}/\texttt{first24}).
For MaxCut, AR is the best returned cut divided by the optimum certified by a classical state-of-art solver (in this case we used CPLEX), averaged over completed instances. For portfolio optimization problem the objective is normalized to the exact feasible range so that AR $=1$ corresponds to the optimal selection. The densification effect motivated in Section~\ref{sec:method}---residual graphs growing denser as variables are eliminated---is probed here from two directions: SK instances which are fully connected from the outset, and larger MaxCut instances, whose residuals densify under recursive reduction. Throughout, beam is evaluated as a decision-side policy layered on fixed-$p$ QAOA, not as a claim that the quantum circuits alone scale with qubit count.

The SK model is a dense Ising problem in which every pair of spins interacts. In the $\pm1$ variant each coupling is either $+1$ or $-1$, whereas in the Gaussian variant coupling strengths are continuous. Because different couplings can prefer mutually incompatible assignments, no spin configuration can generally satisfy every local preference; this conflict is called frustration. In cardinality-constrained portfolio optimization, a binary variable records whether each asset is selected, covariance terms couple pairs of assets, and a cardinality constraint limits how many assets may be chosen. The non-recursive and recursive arms enforce this constraint differently, so the portfolio results are end-to-end comparisons rather than perfectly controlled mixer comparisons.

Small SK instances use exact-statevector simulators, while larger MaxCut graphs use MPS simulators (with reduced bond dimension of 128, and 32 shots) and re-score cuts against CPLEX. Displayed large MaxCut averages use three complete instances, and SK $\pm1$ averages use ten seeds. Non-recursive gold/LR plain and portfolio QAOA arms use a Dicke initial state with an XY mixer~\citep{hadfield2019qaoa} (feasible subspace), whereas recursive gold/LR RQAOA and beam start from $\lvert+\rangle^{\otimes n}$ with a penalty-folded Ising and may require cardinality repair before scoring, so beam AR~$1.000$ cannot be credited to a Dicke prior the recursive arms never receive. Simulation portfolio results (\texttt{first20}/\texttt{first24}) are separate from the portfolio run on \texttt{ibm\_marrakesh} (\texttt{first16}, $\lambda{=}5\times10^{-4}$).

The initial state and mixer differ between the non-recursive and recursive arms, and this difference matters for interpreting the portfolio results. The non-recursive plain QAOA arms (gold and LR) and the portfolio QAOA arm prepare a Dicke initial state and apply an XY mixer~\citep{hadfield2019qaoa}, which confines the evolution to the feasible subspace of fixed-cardinality selections. The recursive arms---gold and LR RQAOA, together with beam---instead start from the uniform superposition $\lvert+\rangle^{\otimes n}$ over a penalty-folded Ising Hamiltonian, and may need a cardinality-repair step to restore feasibility before a cut is scored. The recursive arms therefore never receive the Dicke prior that keeps the plain arms feasible by construction, so beam's AR of $1.000$ on portfolio cannot be attributed to that advantage; it is achieved from the weaker starting point. Finally, the simulated and hardware portfolio experiments are distinct instances and should not be compared directly: the simulation results are reported for \texttt{first20} and \texttt{first24}, whereas the hardware run on \texttt{ibm\_marrakesh} uses \texttt{first16} with penalty weight $\lambda{=}5\times10^{-4}$.

Primary baselines are fixed-schedule gold and LR RQAOA. We always report non-recursive QAOA when the comparison is complete so the evaluation is not recursive-only: on larger MaxCut graphs this is best-of-shots, while on SK it uses an  top-64 amplitude readout and is \emph{not} a fair single-answer baseline (marked $^\ast$). COBYLA RQAOA (angle-optimized recursion) is a strong classical-in-the-loop baseline when it finishes, but for many large-graph COBYLA runs hit walltime or remain pending (completed runs span roughly $5$--$35$\,h). Reported beam results use $B{=}4$, and walltime ablations also report $B{=}6/8$.

\subsection{Simulation results}

\begin{table}[h]
\centering
\setlength{\tabcolsep}{3pt}
\caption{Simulation approximation ratios (AR; $1.000$ is optimal). Results span different $p$, with most beam cells at $p{=}2$ unless noted; means over completed instances. Columns: gold/LR $=$ fixed-schedule recursive QAOA (gold $=$ evolved power-law schedule); QAOA $=$ non-recursive single-circuit baseline; Beam $=$ sign-hedged beam RQAOA (beam width $B{=}4$). Bold: best fair arm in-row (SK QAOA$^\ast$ excluded). Rows marked $^\dagger$ are counterexample rows. Portfolio $\lambda$ omitted from the body: \texttt{first24} uses $\lambda{=}10^{-3}$; \texttt{first20}/\texttt{first16} use $\lambda{=}5\times10^{-4}$. SK Gaussian and portfolio \texttt{first20} use $p{=}1$.}
\label{tab:sim}
\small
\begin{tabular}{@{}lrrrrr@{}}
\toprule
Instance & $n$ & gold & LR & QAOA & Beam \\
\midrule
Erd\H{o}s--R\'enyi ($k{=}2$) & 40 & 0.849 & 0.855 & 0.885 & \textbf{0.990} \\
Erd\H{o}s--R\'enyi ($k{=}2$) & 60 & 0.825 & 0.793 & 0.837 & \textbf{0.984} \\
5-regular MaxCut & 40 & 0.963 & 0.951 & 0.926 & \textbf{0.996} \\
5-regular MaxCut & 50 & 0.950 & 0.960 & 0.908 & \textbf{0.987} \\
5-regular MaxCut & 60 & 0.945 & 0.956 & 0.877$^{\ddagger}$ & \textbf{0.995} \\
4-regular MaxCut$^\dagger$ & 40 & 0.980 & 0.981 & 0.922 & \textbf{0.990} \\
4-regular MaxCut$^\dagger$ & 60 & 0.954 & \textbf{0.987} & 0.843 & 0.948 \\
SK $\pm1$ & 20 & 0.474 & 0.470 & 1.000$^\ast$ & \textbf{0.998} \\
SK $\pm1$ & 24 & 0.532 & 0.463 & 1.000$^\ast$ & \textbf{0.986} \\
SK Gaussian$^\dagger$ & 20 & 0.306 & 0.306 & 1.000$^\ast$ & \textbf{0.983} \\
Portfolio \texttt{first24}$^\dagger$ & 24 & 0.936$^{r}$ & \textbf{1.000}$^{r}$ & 0.978 & \textbf{1.000}$^{r}$ \\
Portfolio \texttt{first20} & 20 & 0.779$^{r}$ & 0.779$^{r}$ & 0.968 & \textbf{0.983}$^{r}$ \\
Portfolio \texttt{first16}$^\dagger$ & 16 & 0.792 & 0.723 & \textbf{1.000} & \textbf{1.000}$^{r}$ \\
\bottomrule
\end{tabular}\\
 Larger MaxCut rows: MPS correlators, mean over 3 instances (except $^\ddagger$ QAOA over $1/3$). SK: exact statevector, 10 seeds (SK $\pm1$ $n{=}24$: gold/LR $9/10$). Counterexample rows: $^\dagger$4-regular $n{=}40$ loses to COBYLA RQAOA $1.000$; 4-regular $n{=}60$ loses to LR $0.987$; SK Gaussian loses to plains $0.989$; \texttt{first24}/\texttt{first16} tie a baseline at $1.000$. $^{r}$cardinality repair. Plain AR on \texttt{first24}: gold/LR plain $0.848/0.850$.
\end{table}

\textbf{Frustrated Sherrington-Kirkpatrick (SK) graphs}
At SK $\pm1$, $n{=}20$, $p{=}2$, gold/LR RQAOA fall to $0.474/0.470$ while beam reaches $0.998$---$2.11\times$/$2.12\times$ ($+110\%$/ $+112\%$ relative). At $n{=}24$ the same pattern holds ($0.532/0.463$ $\to$ $0.986$, about $1.9\times$/$2.1\times$). Fully connected $\pm1$ couplings make locally large $|\langle Z_iZ_j\rangle|$ globally unsafe, so irreversible largest-$|\mu|$ signs fail even when both recursive baselines already use fixed gold/LR schedules; this is a \emph{decision} failure at constant $p$. For realism we also report SK Gaussian $n{=}20$, $p{=}1$, where non-recursive plains reach $0.989$ and beam $0.983$, a near-tie rather than a domination claim. The method's headline is therefore the frustrated $\pm1$ family rather than universal superiority on every spin-glass draw and overall the method demonstrates largest observed improvement in SK graphs across all other COP instances.

\textbf{Larger MaxCut graphs.}
On Erd\H{o}s--R\'enyi $k{=}2$ MaxCut at $n{=}40$, beam $0.990$ beats non-recursive QAOA $0.885$ ($+11.9\%$ relative) and best fixed recursive $0.855$ ($+15.8\%$). At $n{=}60$, beam $0.984$ vs.\ QAOA $0.837$ ($+17.6\%$) and gold $0.825$ ($+19.3\%$). On 5-regular MaxCut at $n{=}40/50/60$, beam reaches $0.996/0.987/0.995$, above QAOA $0.926/0.908/0.877$. We also show the counterexample rows in Table \ref{tab:sim}- on 4-regular $n{=}40$, $p{=}2$, COBYLA RQAOA saturates at $1.000$ while beam is $0.990$; at $n{=}60$, LR RQAOA $0.987$ beats beam $0.948$. Angle search or a lucky fixed schedule can still win when they finish, but COBYLA is the arm that most often hits walltime.

\textbf{Portfolio.}
On \texttt{first24} ($\lambda{=}10^{-3}$, $p{=}2$), beam ties LR RQAOA at $1.000$ after repair, exceeds gold RQAOA $0.936$ ($+6.8\%$ relative), and edges non-recursive QAOA $0.978$ ($+2.2\%$ relative). On \texttt{first20} ($\lambda{=}5\times10^{-4}$, $p{=}1$), beam reaches $0.983$ ahead of QAOA $0.968$ ($+1.5\%$) and well ahead of both recursive schedules at $0.779$ ($+26\%$ relative to gold/LR). On \texttt{first16} instance matched to the run on \texttt{ibm\_marrakesh} run ties QAOA at $1.000$. We couldnt run a full arm suite on the native QOBLIB $n{=}50$ instance because it is not tractable for the exact or recursive methods. Its covariance graph is the complete graph $K_{50}$, so the dense Ising problem has no sparsity to exploit: an exact $2^{50}$ statevector is infeasible, and the MPS correlators do not converge---the bond dimension hits its truncation ceiling, so the approximation cannot be certified against an exact answer. The cost is also prohibitive, with a single correlation call taking tens of thousands of seconds and a full recursive row taking tens of hours. For these reasons we evaluate the first-$n$ ladder (\texttt{first16}/\texttt{first20}/\texttt{first24}) throughout, rather than the full $n{=}50$ instance.

\textbf{Wall-clock cost vs.\ QAOA and COBYLA.}
Table~\ref{tab:wall} reports MPS walltime on the same larger Erd\H{o}s--R\'enyi and regular MaxCut instances. Non-recursive QAOA on Erd\H{o}s--R\'enyi $k{=}2$ at $n{=}40$, $p{=}2$ takes $29.61$\,h, while beam $B{=}4$ finishes in $5.63$\,h ($\approx 5.3\times$ faster) despite more circuit evaluations, because each residual is smaller. Beam $B{=}6/8$ drop further to $\approx 1.5$\,h: the $B{=}4$ runs are serial, whereas $B{=}6/8$ use a process-pool MPS backend that evaluates independent beam children in parallel, so per-eval walltime is not comparable across widths, but the total-time gap vs.\ QAOA is the operational message. COBYLA RQAOA spans roughly $5$--$35$\,h on completed large-graph runs (Table~\ref{tab:wall} includes a $13.48$\,h example), and several larger runs remain pending after hitting walltime, which is why we treat it as an aspirational rather than always-available baseline.
\begin{table}[t]
\centering
\setlength{\tabcolsep}{2.6pt}
\caption{Simulation wall-clock time on larger MaxCut graphs (MPS backend; mean total elapsed). Beam width $B{=}4$ is serial; $B{=}6/8$ use parallel MPS workers. QAOA layer depth $p$ is listed with $n$. ER $=$ Erd\H{o}s--R\'enyi.}
\label{tab:wall}
\small
\begin{tabular}{@{}llrrrrr@{}}
\toprule
Family & $n{,}p$ & gold & LR & COBYLA & Beam$_4$ & QAOA \\
\midrule
ER ($k{=}2$) & 40,2 & 1.27\,h & 1.28\,h & 15.60\,h & 5.63\,h & 29.61\,h \\
ER ($k{=}2$) & 60,1 & 2.54\,h & 2.49\,h & 20.77\,h & 9.47\,h & 17.12\,h \\
4-regular & 40,2 & 36.7\,m & 43.1\,m & 13.48\,h & 2.54\,h & 16.05\,h \\
\bottomrule
\end{tabular}\\
Parallel ablations on ER $n{=}40$, $p{=}2$: Beam$_6$ $1.48$\,h, Beam$_8$ $1.47$\,h (vs.\ serial Beam$_4$ $5.63$\,h). COBYLA pending on several larger cells.
\end{table}

\subsection{Hardware results and resource costs}

These are single-instance demonstrations: the arms are not depth-matched, and we claim neither quantum advantage nor a billed-runtime speedup over classical solvers. Instance selection used a hard post-transpilation two-qubit-gate depth filter (${<}200$) on \texttt{ibm\_kingston} as the primary admission rule, where post-transpilation two-qubit-gate depth is the longest sequence of entangling-gate layers after the circuit is mapped to the device. We ran three families under that gate---Erd\H{o}s--R\'enyi $k{=}2$ MaxCut at $n{=}40$, $p{=}2$ (2-qubit depth $156$); SK $\pm1$ at $n{=}20$, $p{=}2$ (2-qubit depth $184$); and portfolio \texttt{first16} (depth $144$)---and skipped surveyed instances at depth $\ge 200$ (e.g.\ Erd\H{o}s--R\'enyi $n{=}60$, $p{=}2$ at $224$, SK $n{=}24$ at $224$, \texttt{first24} at $224$, Erd\H{o}s--R\'enyi $n{=}40$, $p{=}3$ at $234$); a few depth-passing 5-regular instances ($120$--$186$) couldnt be run for QPU-budget reasons and will be run as next steps.

\begin{figure}[H]
\centering
\includegraphics[width=\columnwidth,height=0.22\textheight,keepaspectratio]{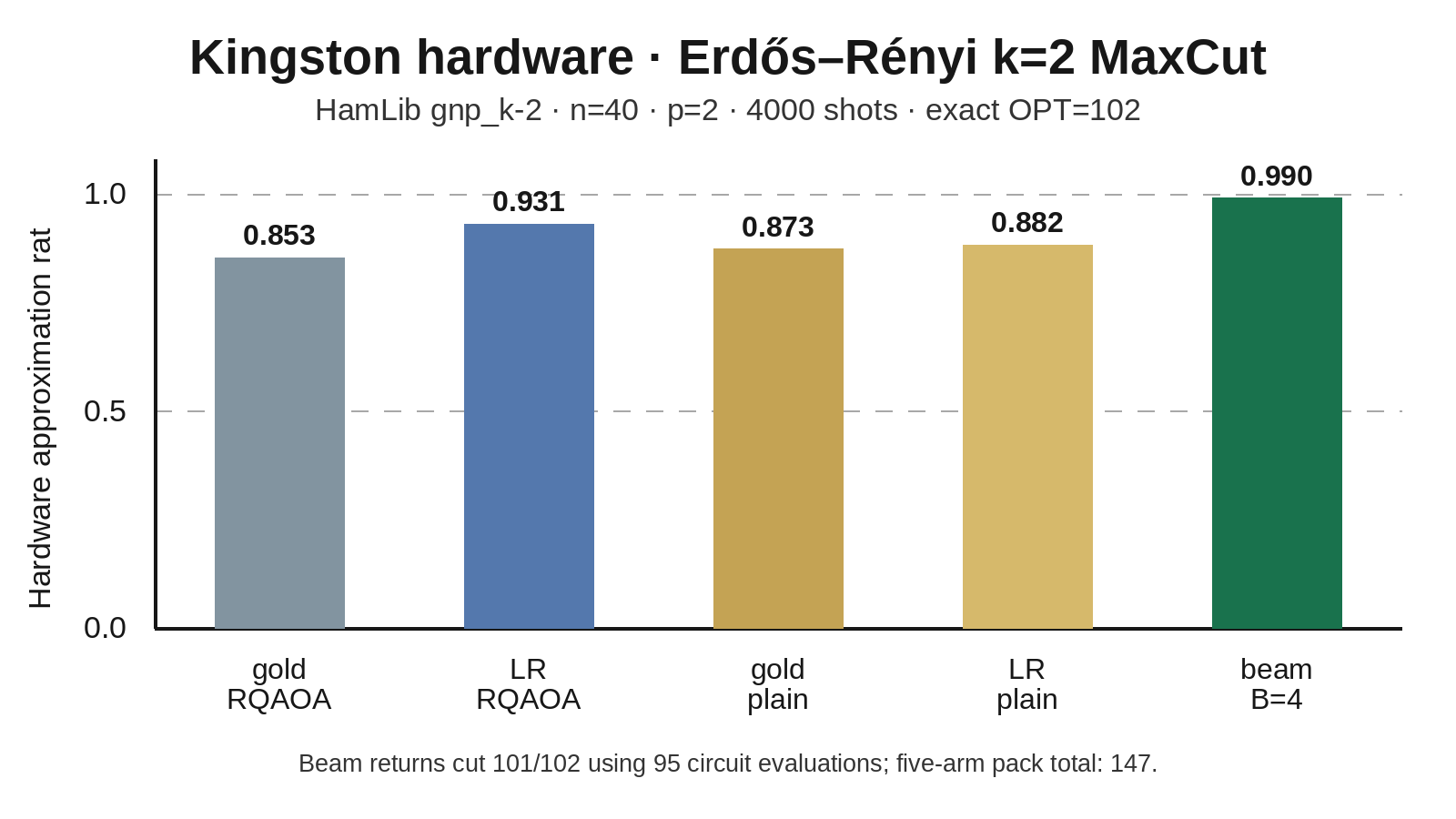}
\caption{Single-instance Erd\H{o}s--R\'enyi $k{=}2$ MaxCut on \texttt{ibm\_kingston} ($n{=}40$, $p{=}2$; 4000 shots/eval). Approximation ratio (AR; $1.000$ optimal): beam reaches $0.990$ and leads all four fixed-schedule baselines. Post-transpile two-qubit-gate depth $156$.}
\label{fig:kingston}
\end{figure}

\begin{figure}[H]
\centering
\scriptsize\makebox[\columnwidth]{\makebox[0.5\columnwidth]{SK $\pm1$ on \texttt{ibm\_marrakesh}}\makebox[0.5\columnwidth]{Portfolio on \texttt{ibm\_marrakesh}}}\\[1pt]
\includegraphics[width=\linewidth,trim=0 0 0 23pt,clip]{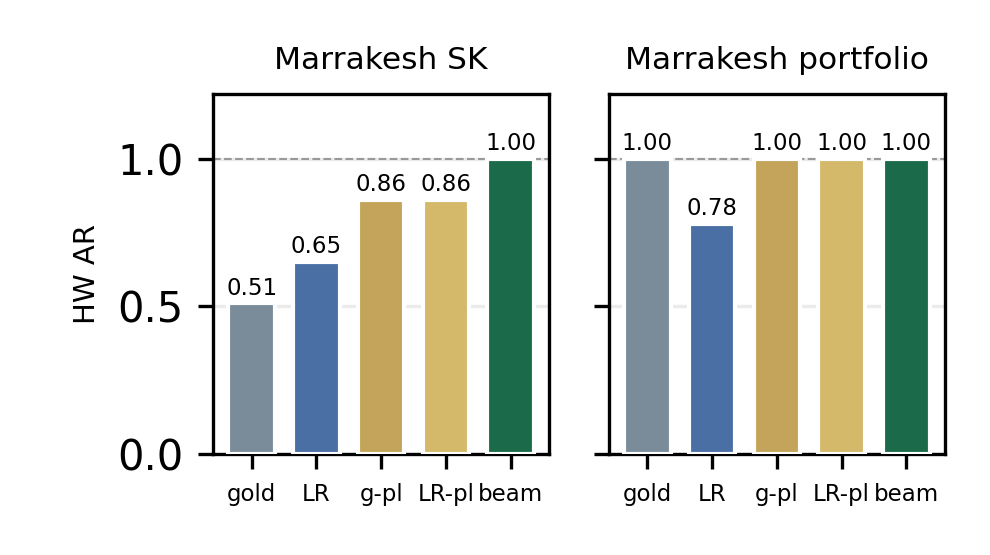}
\caption{Single-instance SK $\pm1$ and portfolio \texttt{first16} on \texttt{ibm\_marrakesh} at $p{=}2$, $4000$ shots/eval. Approximation ratio (AR): beam reaches the SK optimum; portfolio ties at AR $1.000$.}
\label{fig:hw}
\end{figure}

For Erd\H{o}s--R\'enyi $k{=}2$ MaxCut ($n{=}40$, $p{=}2$) on \texttt{ibm\_kingston}, beam reaches AR $0.990$ versus $0.853$ (gold RQAOA; $+16.1\%$ relative), $0.931$ (LR RQAOA; $+6.3\%$), $0.873$ (gold plain), and $0.882$ (LR plain) (Fig.~\ref{fig:kingston}). For SK $\pm1$ ($n{=}20$, $p{=}2$) on \texttt{ibm\_marrakesh}, beam reaches the optimum ($43/43$, AR $1.000$) vs.\ gold/LR RQAOA $0.512/0.651$ ($\approx 2.0\times$/$1.5\times$ improvement) and with gold-plain $0.860$ ($+16.3\%$, $1.16\times$). For portfolio \texttt{first16} on \texttt{ibm\_marrakesh}, beam/gold/plains reach $1.000$ while LR RQAOA reaches $0.783$ (Fig.~\ref{fig:hw}).

Post-transpile depths for the admitted circuits are in Table~\ref{tab:hwdepth}. Recorded QPU seconds are in Table~\ref{tab:qpu} (pack $399$\,s recorded; null usage metadata on some jobs makes this a lower bound). QPU seconds measure recorded device-processing time rather than total elapsed wall time, and the plains here are RZZ cost-Ising circuits rather than Dicke$+$XY.

\begin{table}[H]
\centering
\setlength{\tabcolsep}{4pt}
\caption{Post-transpilation two-qubit-gate depth on \texttt{ibm\_kingston} (all ${<}200$). Depth $p{=}2$ for every row. Two-qubit-gate depth $=$ longest entangling-gate layer sequence after device mapping.}
\label{tab:hwdepth}
\small
\begin{tabular}{@{}lrr@{}}
\toprule
Circuit family & $n$ & two-qubit-gate depth \\
\midrule
Erd\H{o}s--R\'enyi ($k{=}2$) MaxCut & 40 & $156$ \\
SK $\pm1$ & 20 & $184$ \\
Portfolio \texttt{first16} & 16 & $144$ \\
\bottomrule
\end{tabular}
\end{table}

\begin{table}[H]
\centering
\setlength{\tabcolsep}{3.5pt}
\caption{Billed IBM Runtime QPU seconds (device-processing time, not wall time) for single-instance runs on \texttt{ibm\_marrakesh} ($4000$ shots/eval). Columns: Beam; gold/LR RQAOA; gold/LR plain QAOA.}
\label{tab:qpu}
\small
\begin{tabular}{@{}lrrrrr@{}}
\toprule
Instance & Beam & gold & LR & gold plain & LR plain \\
\midrule
SK $\pm1$ & $153$\,s & $41$\,s & $37$\,s & $4$\,s & $4$\,s \\
Portfolio \texttt{first16} & $96$\,s & $32$\,s & $29$\,s & $3$\,s & $0$\,s \\
\bottomrule
\end{tabular}
\end{table}

\section{Discussion}
\label{sec:discussion}

Parameter selection and recursive decision-making limit fixed-$p$ recursion in different ways, and our two contributions target them separately. A fixed schedule attacks the first: gold removes instance-wise angle search and can raise success probability at a given $p$ (Fig.~\ref{fig:gold-transfer}). But a better schedule only sharpens the correlators a solver is handed; it cannot repair a wrong decision made from them. That is the second bottleneck, and it is where beam acts: given imperfect correlators at the chosen $p$, standard RQAOA freezes a single sign irrevocably, whereas beam retains both signs and keeps $B{=}4$ residuals at the same $p$. The two levers are therefore complementary rather than competing---gold supplies better inputs to the decision, beam makes the decision recoverable.

The failure beam repairs is not an artifact of our instances but a structural property of irreversible recursion. Concurrent theoretical work by Hadfield et al.~\citep{hadfield2026iterative} proves this for an infinite family of unweighted MaxCut instances: level-1 QAOA drives recursive reduction to strictly prefer a first elimination that excludes every optimal solution, capping that branch at approximation ratio $2/3+o(1)$, while a single backtrack to the complementary branch recovers an optimum from the same quantum information. Our SK $\pm1$ result is an empirical, hardware-validated instance of exactly this phenomenon: at AR $0.47$ the correlators were informative enough to solve the problem, yet the irrevocable commitment discarded the optimum, and hedging recovers it to $0.998$. Sign hedging and their reversible tree are two realizations of one fix---ours bounded to width $B$ at fixed $p$ so residual circuits stay runnable on current quantum devices, theirs extending to full certification when the classical search budget permits.

  The experiments isolate the decision as the controlling factor. SK makes this cleanest: both recursive baselines already use the fixed gold/LR schedules, so their collapse to $\approx 0.47$ cannot be blamed on parameters---only the irreversible sign differs from beam, and only beam recovers near-optimal AR. The larger MaxCut graphs show the effect is not special to SK: the same ranking holds through $n{=}60$ without deeper circuits, and because each residual is smaller, beam reaches it in a fraction of non-recursive QAOA's walltime on the same backend. Portfolio adds a third angle, where beam stays competitive even though the recursive arms never receive the feasibility-preserving Dicke$+$XY prior the non-recursive plains enjoy. Across all three families the gains come from making the elimination recoverable, not from increasing $p$. Beam is thus a decision-side optimization on top of fixed-$p$ QAOA/RQAOA; its relevance at larger qubit counts is conditional on residual circuits remaining runnable, and the SK densification regime is our reason to expect the same policy to transfer as future hardware admits those circuits.

Three caveats bound these results. First, the gold schedule was discovered by CliffSearch's evolutionary, LLM-guided formula search~\citep{cliffsearch}; we treat that only as the offline discovery mechanism for a fixed ramp, not as part of the runtime algorithm. Second, the hardware ARs on \texttt{ibm\_kingston} and \texttt{ibm\_marrakesh} are raw Sampler outcomes at $4000$ shots, with no error mitigation, dynamical decoupling beyond backend defaults, or probabilistic error cancellation---so they are a floor, not a tuned result. Third, beam's extra cost is classical and parallelizable: serial $B{=}4$ already runs faster than non-recursive QAOA in simulation, and parallel $B{=}6/8$ process-pool runs cut walltime by a further $\approx 3$--$4\times$, but on hardware this parallelism would require issuing independent beam children across multiple QPUs with matched noise profiles so that pruning stays statistically fair. A controlled follow-on should therefore compare raw against error-mitigated hardware ARs, account wall- and QPU-time for matched-noise multi-QPU beam children, and run a factorial gold/LR $\times$ beam sweep as $p$ grows on larger graphs where residual densification intensifies.

\FloatBarrier
\section{Conclusion}
\label{sec:conclusion}

We introduced sign-hedged beam RQAOA, a bounded decision-side search that makes recursive elimination reversible at fixed $p$. Its premise is that the binding constraint on hard instances is not circuit depth but the irrevocable commitment to an elimination sign; by retaining both signs and pruning to the $B{=}4$ most promising residuals, beam recovers solutions that single-path recursion discards. The clearest evidence is frustrated SK $\pm1$, where standard recursion collapses to AR $\approx 0.47$ while beam reaches $0.998$ in simulation and the optimum on hardware, with the advantage carrying across larger MaxCut and portfolio instances on both simulator and device. Because beam spends only classical effort rather than deeper circuits, it is the natural partner to a fixed schedule and is positioned to scale as future hardware makes larger residual circuits runnable---the direction we see as most promising for pushing recursive quantum optimization beyond the sizes reachable today.

\section*{Disclosure}

The authors used large language models (LLMs) as computational assistants during the research process, including for brainstorming alternative approaches, critiquing and validating implementation ideas, and accelerating software development through code generation. All generated content was critically evaluated, modified where necessary, experimentally verified, and integrated by the authors. The conception of the research, scientific contributions, interpretation of results, and conclusions are solely those of the authors. Implementation code will be made available upon reasonable request, and will be released publicly upon publication.

\bibliographystyle{plainnat}
\balance
\bibliography{refs}
\end{document}